\documentclass{ere04}

\usepackage{graphicx}

\usepackage{amssymb}       

\begin{document}

\title*{Holographic Cosmology and Uncertainty Relation}
\author{M. A. Per\and A. Segu\'\i }
\institute{Departamento de F\'\i sica Te\'orica. Facultad de Ciencias. Universidad de
Zaragoza. \texttt{segui@unizar.es}}
\maketitle

It is believed that a primary principle of the theory of quantum gravity is the Holographic
Principle according to which a physical system can be described only by degrees of freedom
living on its boundary. The generalized covariant formulation of the principle considers the
entropy content on truncated \emph{light-sheets}: light-like hypersurfaces of non-positive
expansion orthogonally generated from a boundary. When we construct the truncated
light-sheets for cosmological observers we find a general expression for the minimum cosmic
time interval from the apparent horizon to verify the holographic prescription; this minimum
time is related to quantum effects involved in the entropy computation. Finally, we arrive to
the uncertainty relation from the Holographic Principle, which suggests a deep connection
between general covariance, entropy bounds and quantum mechanics.

\section{Introduction}
\begin{figure}[!h]
\begin{center}
\includegraphics[width=12cm]{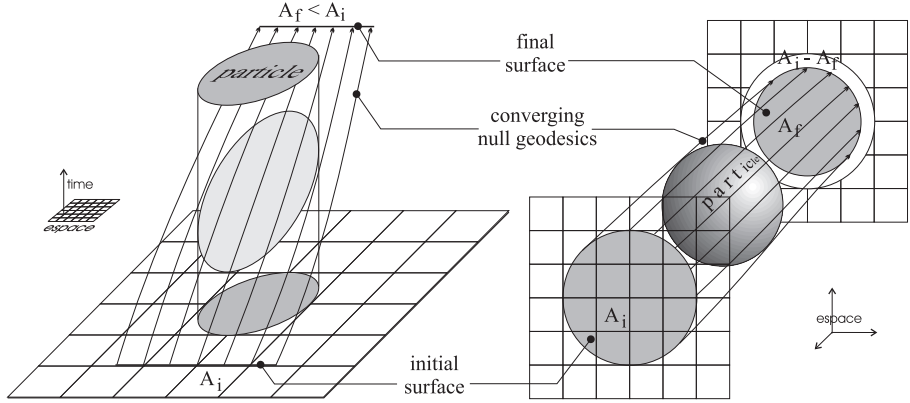}
\end{center}
\caption{{\small Reduction of the cross sectional area of a LS crossing a fundamental
particle. On the left we can see a 2+1 dimensional representation and on the right a purely
spatial one. The deflection angle gravitationally induced by the particle is proportional to
$M/R$; the radius reduction is then $\Delta R \sim \frac{M}{R}\ c \Delta t$ and thus, the
difference of the boundary areas is, approximately: $\Delta A \sim 2 \pi \ M \ \Delta t $.
Assuming that the entropy associated to a fundamental particle is $S \sim 1$, when we apply
the GCEB (\ref{slb}) we arrive to the energy-time uncertainty relation $\Delta E \Delta t
\gtrsim \hbar$.}}\label{part}
\end{figure}

One of the most promising ideas that emerged in theoretical physics during the last decade
was the Holographic Principle proposed by 't Hooft and Susskind \cite{tHooft,Susskind}; it
appears to be a new guiding paradigm for the true understanding of quantum gravity theories.
Basically, the Holographic Principle states that the fundamental degrees of freedom of a
physical system
are bound by its surface area in Planck units.\\
The covariant formulation of the Principle is proposed by Bousso \cite{Bousso1,Bousso2} who
considers the entropy content on the \emph{light-sheets} (LS): light-like hypersurfaces of
non positive expansion orthogonally generated from the boundary $B$ of the physical system.
Later, a more refined version was given by Flanagan, Marolf and Wald \cite{FMW,BFM,ST} who
proposed the Generalized Covariant Entropy Bound (GCEB): the entropy content of a LS
\emph{truncated} by another boundary $B'$ must be bound by one quarter of the difference of
the boundary areas
\begin{equation}\label{slb}
  S[L(B,B')] \leq \frac{A-A'}{4} \, .
\end{equation}
From this basic principle it is easy to deduce the Generalized Second Law of Thermodynamics
and the intriguing Bekenstein entropic limit $S \leq 2 \pi \ Energy \times Radius $
\cite{BoussoBek1,BoussoBek2}. Also, from the GCEB we can show a heuristic connection between
the Holographic Principle and the uncertainty relations from a microscopic point of view
proposed by Bousso \cite{BoussoUncert}: the basic idea is to apply the GCEB to a fundamental
particle (Fig.~\ref{part}). Another connection between the HP and the uncertainty relations
has been shown by Jack Ng working at a very fundamental level related to the space-time foam
\cite{Ng1,Ng2}. We also mention another previous works in that line
\cite{preur1,preur2,preur3}.

~\\
But our work has followed a different line: The GCEB is the strongest formulation of the
entropic bound and by imposing it on different physical systems we look for further insights
on the nature of the Holographic Principle and the theory of quantum gravity; particularly,
we impose the GCEB on isotropic cosmological scenarios. We will find that the verification of
the GCEB needs a restriction very similar to the energy-time uncertainty relation. At first
sight, this restriction might be considered a drawback; however, we consider that we are in
front of a possible confirmation of the fundamental nature of the Holographic Principle.

\section{The Apparent Horizon}

\begin{figure}[!hbt]
~\\[-1.8cm]
\begin{center}
\includegraphics[height=6cm, width=14.5cm]{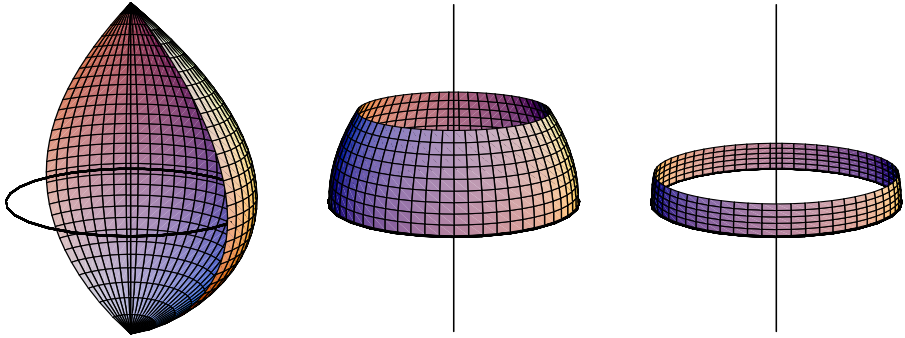}
\end{center}
~\\[-2.0cm]
\caption{{\small Proper distance-time representation of a past light cone, and two truncated
light sheets traced from the apparent horizon; on the right we see how a very little time
interval might violate the GCEB: the difference between the bounding areas (circles in the
figure) tend to zero.}}\label{nul}
\end{figure}

For an easy area calculation, and also in order to improve the graphic representation, we
will work with proper distances $D$ rather than coordinate distances $r$. In this context it
is easy to define the proper distance from the fiducial observer located at $r=0$ to an event
with coordinate distance $r$, at the same cosmological time $t$. We define the proper
distance integrating the spatial part of the Robertson Walker metric (only the radial part is
significant here).
\begin{equation}\label{metric}
  ds^{2}=-dt^{2}+R^{2}(t) \Big( \frac{dr^{2}}{1-k r^{2}}+r^{2}d\Omega^{2}_{n-1} \Big)
  \quad \Rightarrow \quad D(r,t) = R(t) \int ^{r}_{0} \frac{dr'}{\sqrt{1-k \, r'^{2}}} \, .
\end{equation}
Thus we rewrite the significant part of the Robertson-Walker metric using now proper
distances
\begin{equation}
  ds^{2}=-dt^{2}+ \big( dD - D\frac{\dot{R}}{R} \, dt \big) ^{2} \, .
\end{equation}
Equating it to zero we get a very descriptive equation for the null geodesics: it is a simple
algebraic sum of velocities. The velocity of the null geodesic with respect to the observer
at $D=0$ is the sum of the recessional velocity of the cosmic fluid there (Hubble law) plus
the velocity of the null geodesic with respect to that fluid, which is always $\pm c$
\cite{Ellis,Kiang,Kinematic} .
\begin{equation}\label{sumavel}
  \dot{D}=HD \pm 1 \quad \Rightarrow \quad V_{null \ \ \, \atop geodes.}=v_{cosmic \ \ \atop recession}  \pm c \, .
\end{equation}
Therefore, when we construct the LS associated with the past light cone of a fiducial
cosmological observer, we found that locally, the past light cone must recede with an angle
of 45 degrees, like in the Minkowsky space-time; on the other hand, at the Big Bang all the
distances must contract to zero. Thus, using space-time proper coordinates, we get a typical
drop-shaped figure (Fig.~\ref{nul}, left).

The cross sectional area of the past light cone reaches a maximum at the so called Apparent
Horizon (AH). In spatially flat n+1 dimensional universes the maximum past light cone cross
sectional area $A(t)=\omega _{n-1}D(t) ^{n-1}$ coincides with the maximum proper distance $D$
from the fiducial observer to the ingoing null geodesics; then, equating (\ref{sumavel}) to
zero (taking the minus sign because we must consider only ingoing null geodesics) we found
that the AH is located at the Hubble distance $H^{-1}$.\\
But for a non-spatially flat universe we can not use the proper distance $D$ like a
\emph{radius} for the area calculation. In such a case, integrating the metric (\ref{metric})
for constant $r$ and $t$ we arrive to a suitable definition of the \emph{areal radius}
$\textsf{R}$
\begin{equation}\label{Ar}
  A(t,r)=\omega_{n-1}\big[ R(t) r  \big] ^{n-1} \quad  \Rightarrow \quad
  \textsf{R} \equiv R(t) r \, .
\end{equation}
Thus, the general expression for the location of the AH is very similar to the geometric term
of the Friedmann equation:
\begin{equation}\label{rah}
  \textsf{R}_{AH}(t)=  \Big(  \frac{\dot{R}^{2}}{R^{2}} +\frac{k}{R^{2}} \Big)
  ^{-\frac{1}{2}} \, .
\end{equation}

The AHs of a fiducial observer on a cosmological model have been proposed by Bousso like
preferred screens for the holographic codification of all the information of the universe
\cite{Bousso1,Bousso2}. Actually, they are not horizons but they are very special surfaces:
they are connected to three LSs; two of them form every past light cone (Fig.~\ref{nul},
left).

\section{Minimum cosmic time interval}

We compute the entropy content of a truncated light sheet integrating the entropy density
$s(t)$ (not strictly \emph{local}) over the corresponding null hypersurface defined by
$\textsf{R}(t)$. The GCEB (\ref{slb}) says that
\begin{equation}
  S_{LS}(t_{i},t_{f}) = \int _{t_{i}} ^{t_{f}} \omega_{n-1} \textsf{R}(t) ^{n-1} s(t) dt =
  \int _{t_{i}} ^{t_{f}} A(t) s(t) dt \, \leq \, \frac{| A(t_{i})-A(t_{f}) |}{4} \, .
\end{equation}
\begin{figure}[!htb]
\begin{center}
\includegraphics[height=4cm, width=12cm]{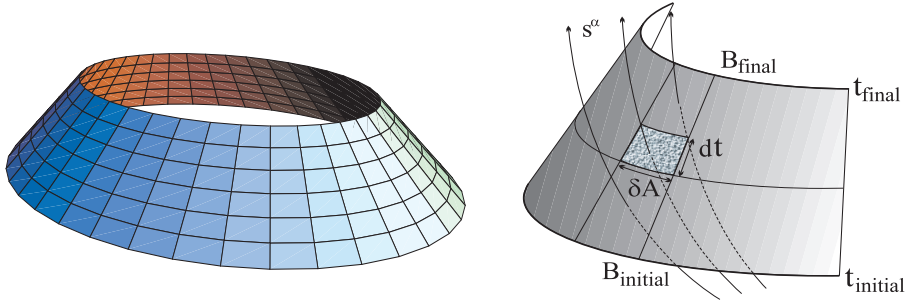}
\end{center}
\caption{{\small Entropy content of a truncated light sheet: integration of the entropy
density.}} \label{gcebf}
\end{figure}
But, in the Fig.~\ref{nul} we can see that near the AH the difference of the boundary areas
truncating the LS may be arbitrarily small, and then the GCEB is violated. Effectively,
taking the Taylor expansion near the AH it is easy to find that the entropy content on the
truncated LS grows linearly with time interval $\Delta t=t-t_{AH}$ but the difference between
the boundary areas grows with the squared time interval (because $\dot{A}(t_{AH})=0$). In a
n+1 dimensional universe we have the following expressions for the entropy content on the LS
generated from the AH and the area available:
\begin{equation}
  S_{LS}(t_{AH},t) \simeq A(t_{AH}) s(t_{AH})  \Delta t \, \quad \& \quad
  \frac{\Delta A (t_{AH},t)}{4} \simeq  -\frac{1}{8}\ddot{A}(t_{AH}) ( \Delta t ) ^{2} \, .
\end{equation}
In order to respect the GCEB we must limit the minimum separation in cosmic time between the
boundary areas on the truncated LS; thus, we have to wait a minimum time interval $\Delta t
_{min}$ so that the squared time interval $(\Delta t ) ^{2}$ term dominates over the linear
one $\Delta t$ and the GCEB would be applied. Assuming that the minimum time interval
verifies $\Delta t _{min} \ll t_{AH}$, we can obtain its approximate value from the Taylor
expansions saturating the bound (\ref{slb}):
\begin{equation}
  S_{LS}(t_{AH},t)< \frac{1}{4}\Delta A (t_{AH},t) \quad \Rightarrow \quad
  \Delta t > \Delta t_{min}=-8 \, \frac{A}{\ddot{A}}\, \bigg| _{AH} s(t)  \, .
\end{equation}
Now we are going to simplify this expression. At the AH, $\dot{A}=\dot{\textsf{R}}=0$; so, we
have the relation $\dot{R}r=-R\dot{r}$, which enable us to transform
\begin{equation}
  \frac{\ddot{A}}{A} \, \bigg| _{AH}=
  (n-1) \frac{\ddot{\textsf{R}}}{\textsf{R}} \, \bigg| _{AH}=
  (n-1) \bigg[ \frac{\ddot{R}}{R}-\Big( \frac{\dot{R}}{R} \Big) ^{2}-\frac{k}{R^{2}} \bigg] \, .
\end{equation}
Thus we have an interesting expression for the minimum cosmic time interval
\begin{equation}\label{dtmi}
  \Delta t _{min} = \frac{-\frac{8}{n-1}s}{\dot{H}-\frac{k}{R^{2}}} \, ,
\end{equation}
where the denominator is exactly the geometrical term of one of the Friedman equations. So we
introduce it to find a simple (dimension independent) dynamical solution
\begin{equation}\label{dtmf}
  \Delta t _{min} = \frac{1}{\pi } \, \frac{s}{\rho +p} = \frac{1}{\pi (1+ \omega )} \, \frac{s}{\rho} \, .
\end{equation}
This general result stresses the deep relation between entropy bounds and the entropy-mass
ratio for general systems. Anyway, we have assumed that $\Delta t _{min} \ll t_{AH}$; so, it
is important to know the behavior of this relation as the universe evolves: assuming now
$\omega$ constant and adiabatic expansion we have $s \sim R^{-n}$ and $\rho \sim R^{-n(1 +
\omega)}$, and then introducing the temporal evolution for flat universes $R \sim
t^{\frac{2}{n(1+ \omega)}}$ we arrive to
\begin{equation}
  \Delta t_{min} \sim R^{n \omega} \quad \Rightarrow \quad
  \frac{\Delta t_{min}}{t_{AH}} \sim \big( t_{AH} \big) ^{\frac{\omega -1}{\omega +1}} \ .
\end{equation}
We find that the size of $\Delta t_{min}$ relative to the cosmic time scale $t_{AH}$ is a
decreasing function for  $\omega < 1$, according to the Fischler-Susskind limit \cite{FS}.

\section{Quantum conclusions}

\begin{itemize}
\item Let us apply the relation (\ref{dtmf}) to the mass-entropy traversing the truncated LS
from the AH: basically we have a restriction on the temporal extension of the light sheet
\begin{equation}
  \Delta t \geq \Delta t_{min} \simeq \frac{s}{\rho}=\frac{S_{LS}}{M_{LS}}  \, .
\end{equation}
If we consider only the necessary volume for the existence of only 1 bit of information, we
denote the necessary mass-energy for this minimum information $M_{1}$, and then, making
explicit the presence of the Planck constant (here $1= \hbar/c^{2}$), we arrive to:
\begin{equation}
  \Delta t M_{1} \gtrsim \hbar \, .
\end{equation}
This familiar equation states that the location of 1 bit in a very little time interval is
only possible if the bit carries enough mass-energy: this mass is necessary for the LS area
reduction, so that the holographic codification of the bit can take place. Conversely, if 
the mass involved in the bit is very little, a big time interval will be necessary for an
enough LS area reduction.\\

\item Now, thinking about the physical process required for the existence of 1 bit, we need
to change a quantum state to an orthogonal state. According to the Margolus-Levitin theorem
\cite{ML}, being $E_{1}$ the average energy of the quantum states, the minimum time required
for this evolution verifies
\begin{equation}
  \Delta t E_{1} \geq \pi \hbar / 2 \, ,
\end{equation}
according to our previous result. This theorem impose a physical limit on the maximum speed
on information \emph{processing}; but at first sight the Holographic Principle impose a
physical limit on information \emph{storage}. We think that the GCEB formulation, evaluating
the entropy content on null sections (they evolve also in the time direction), provide an
unified restriction. \\

\item We can guess a microscopic explanation of the restriction encountered in the GCEB: if the
expansion is adiabatic, the restricted time (\ref{dtmf}) adopts the form
\begin{equation}
  s=\frac{\rho + p}{T} \quad \Rightarrow \quad \Delta t_{min} =
  \frac{1}{\pi T_{AH}} \simeq \lambda_{max}
\end{equation}
where $T_{AH}$ is the temperature of the fluid on the AH; its inverse will be a measure of
the typical wavelength of the quanta that carries the entropy. So, when we count the entropy
on the LS, it is natural to consider only those modes whose wavelength is smaller than the
size of the LS; that is $\lambda < \lambda_{max}$ \cite{BFM}. In \cite{cures} we show that
the entropy density, restricted by this kind of IR cutoff,
always yield an entropy content according to the GCEB.\\

\item Finally, we would like to stress the significant emergence of the Friedmann equations in this
context. The relations (\ref{rah}) and (\ref{dtmi}) allow us to rewrite the Friedman
equations like a relation between dynamical terms and the evolution of the AH area; using the
areal radius $\textsf{R}$ (\ref{Ar}) we have:
\begin{equation}
  \textsf{R}_{AH}^{-2} \, = \, \frac{16 \pi G}{n(n-1)} \, \rho \quad \quad \&  \quad \quad
  \frac{\ddot{\textsf{R}}_{AH}}{\textsf{R}_{AH}}  \, = \,  -\frac{8 \pi G}{n-1} \,
  (\rho + p) \, .
\end{equation}
For our traditional vision of the world, it is more suitable to know the evolution of the
scale factor. But, the peculiar form that now we have obtained perhaps suggest an alternative
vision related to the horizon area; an \emph{holographic} vision. In any case, we think that
we are in front of a deep connection between general covariance, entropy bounds and quantum
mechanics; this might be the way to the ultimate theory.

\end{itemize}

\begin{figure}[!hbt]
~\\[-0.6cm]
\begin{center}
\includegraphics[height=6.0cm, width=12cm]{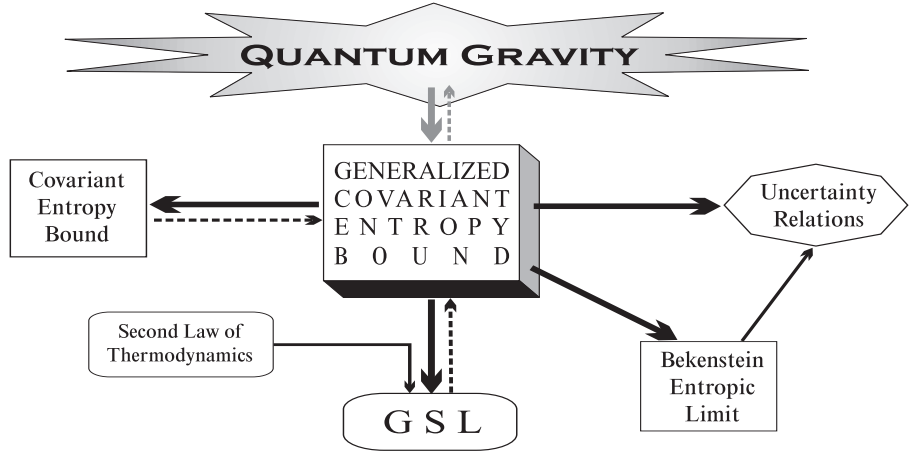}
\end{center}
~\\[-1.0cm]
\caption{{\small Connections between entropic limits and fundamental laws. The dashed arrows
are used when a law was useful for the formulation of a more general law; this subsequent law
generally enable the logical deduction of the previous one (continuous arrows). Obviously,
the connection with the theory of quantum gravity is not clear yet.}}
\end{figure}

\section*{Acknowledgements}

We would like to thank the organizers for this friendly and productive meeting. The present
work has been partially supported by MCYT (Spain) under grant FPA2003-02948



\begin{thebibliography}{99.}


\bibitem{tHooft} G.~'tHooft: ``Dimensional reduction in quantum gravity''.
    In \emph{Salanfestschrift} pp. 284-296, ed. by A. Alo, J. Ellis,
    S. Randjbar-Daemi, World Scientific Co, Singapore, 1993 [gr-qc/9310006].

\bibitem{Susskind} L.~Susskind, J. Math. Phys. 36 (1995) 6377 [hep-th/9409089].

\bibitem{Bousso1} R.~Bousso, Class. Quant. Grav. 17 (2000) 997 [hep-th/9911002].

\bibitem{Bousso2} R.~Bousso, Rev. Mod. Phys. 74 (2002) 825-874 [hep-th/0203101].

\bibitem{FMW} E. E. Flanagan, D. Marolf and R. M. Wald, Phys. Rev. D62 (2000) 084035 [hep-th/9908070].

\bibitem{BFM} R. Bousso, � Flanagan and D. Marolf, Phys. Rev. D 68, 064001 (2003) [hep-th/0305149].

\bibitem{ST} A. Strominger and D. Thompson, Phys.Rev. D70, 044007 (2004) [hep-th/0303067].

\bibitem{BoussoBek1} R. Bousso, Phys. Rev. Lett. 90 (2003) 121302 [hep-th/0210295].

\bibitem{BoussoBek2} R. Bousso, JHEP 0402, 025 (2004) [hep-th/0310148].

\bibitem{BoussoUncert} R. Bousso, JHEP 0405 (2004) 050 [hep-th/0402058].

\bibitem{Ng1} Y. J. Ng, Mod.Phys.Lett. A18 (2003) 1073-1098 [gr-qc/0305019]15.

\bibitem{Ng2} Y. J. Ng, Phys. Rev. Lett. 86, 2946-2949 (2001) [gr-qc/0006105]6.

\bibitem{preur1} M. G. Ivanov and I. V. Volovich, Entropy 3 (2001) 66-75 [gr-qc/9908047].

\bibitem{preur2} F. Scardigli and R. Casadio, Class.Quant.Grav. 20, 3915-3926 (2003) [hep-th/0307174]5.

\bibitem{preur3} D. Minic, Phys.Lett. B442 (1998) 102-108 [hep-th/9808035]16.

\bibitem{Ellis} G.~F.~R.~Ellis and T.~Rothman, Am. J. Phys. 61 (1993) 883.

\bibitem{Kiang} T. Kiang, Chin. Astron. Astrophys. 27/3: 247-253 (2003) [astro-ph/0305518].

\bibitem{Kinematic} L.~J.~Boya, M.~A.~Per and A.~J.~Segu\'\i, Phys. Rev. D 66, 064009 (2002) [gr-qc/0203074].

\bibitem{FS} W. Fischler and L. Susskind, SU-ITP-98-39, UTTG-06-98 [hep-th/9806039].

\bibitem{ML}  N. Margolus and L. B. Levitin, Physica D120 (1998) 188-195 [quant-ph/9710043].

\bibitem{cures} M.~A.~Per and A.~J.~Segui:
    ``Problems and cures (partial) for holographic cosmology''.
    In \emph{Proceedings of "String Phenomenology 2003"}, World Scientific,
    Durham, 2003 [hep-th/0310028].

\end{thebibliography}
\end{document}